
\magnification=\magstep1
\nopagenumbers
\baselineskip=12pt                      
\hsize=16.5truecm
\vsize=24truecm
\font\title = ambx10 scaled 1440        

\font\small = cmr7
\newcount\ftnumber
\def\ft#1{\global\advance\ftnumber by 1
          {\baselineskip 12pt
           \footnote{$^{\the\ftnumber}$}{#1}}}
\newcount\eqnumber
\def\equ(#1){\global\advance\eqnumber by 1
    \expandafter\xdef\csname !#1\endcsname{\the\eqnumber}
    \eqno(\the\eqnumber)}
\def\(#1){(\csname !#1\endcsname)}
\def\fr#1/#2{{\textstyle{#1\over#2}}}

\def\boxit#1{\vbox{\hrule\hbox{\vrule\kern3pt         
     \vbox{\kern3pt#1\kern3pt}\kern3pt\vrule}\hrule}} 
\def\ie{{\it i.e.\/}$\,\,\,$}

\def\={\equiv}

\def\b#1{{\bf b}^{(#1)}}

\def\bi{{\bf b}^{(i)}}
\def\bj{{\bf b}^{(j)}}

\def\ai{{\bf a}^{(i)}}

\def\rv{{\bf r}}
\def\kv{{\bf k}}
\def\cv{{\bf c}}
\def\cvp{{\bf c'}}
\def\cvs{{\bf c_s}}
\def\dv{{\bf d}}
\def\bv{{\bf b}}
\def\av{{\bf a}}

\def\P#1#2{\Phi_{#1}(#2)}
\def\rk#1{\rho({\bf k}_#1)}
\def\rhok{\rho({\bf k})}
\def\An{A({\bf k}_1\ldots{\bf k}_n)}
\def\A#1n{A({#1\bf k}_1\ldots{#1\bf k}_n)}
\def\ep{\epsilon}

\def\tt{\thinspace\thinspace}

\def\tt{\thinspace\thinspace}

\def\ab{$\fr2/3\av + \fr1/3\bv$}
\def\hp{\hskip0.2truein}

\centerline{{\title Crystallography Without Periodicity}}
\bigskip
\centerline{N. David Mermin}
\centerline{Laboratory of Atomic and Solid State Physics}
\centerline{Cornell University, Ithaca, NY 14853-2501}
\bigskip

\midinsert
\narrower\narrower

{\it Abstract.\/} A condition of physical stability, applied to materials
with sharp Bragg peaks, leads to rules for a unified approach to their
classification that make no use of periodicity, applying equally well to
either periodic or quasiperiodic materials.  Three-dimensional geometric
intuition is the primary analytical tool for extracting the classification
from the rules.  Recognizing that periodicity is not required to formulate
crystallography removes the need for an algebraic formalization of that
intuition in a higher dimensional superspace, where the material can be
embedded in a periodic supercrystal.  Selected examples illustrate the
simplicity of the resulting description of ordinary crystals, quasicrystals,
modulated crystals, intergrowth crystals, or modulated quasicrystals.

\endinsert

\bigskip
\centerline{{\bf I. Introduction}}
\medskip

Four years ago at the XVII$^{{\small th}}$ manifestation of this
International Colloquium I described how to extend the classification scheme
of crystallography to positionally ordered materials\ft{Mermin (1989).} with
crystallographically forbidden point groups---the then newly discovered
quasicrystals---without making a detour into higher dimensional spacegroups,
as in the widely used ``superspace'' formulation.\ft{For a recent exposition
of the superspace approach see Janssen (1992).} I am back today, with
variations on the same theme because several things have happened:

1. Our approach has been refined\ft{The point of view I describe here was
developed during an extended collaboration with Daniel Rokhsar and David
Wright.  We were subsequently joined by David Rabson.  For a recent
formulation and references to earlier work see Mermin (1992) or Rabson
(1991).  The approach to modulated crystals described below was worked out in
collaboration with Ron Lifshitz.  A preliminary report (focusing on Bravais
classes) can be found in Mermin (1992b) and at greater length in Mermin
(1992c).} to give a compact and unified derivation of the 230 space groups
for periodic crystals, together with all the analogous quasicrystallographic
categories on so-called ``standard lattices''.\ft{All lattices compatible
with point groups of rotational symmetry less than 23-fold are standard.}

2. Our approach has been applied to arbitrary types of quasiperiodic
ordering: modulated crystals and intergrowth crystals as well as
quasicrystals.  We believe the conventional superspace crystallography of
modulated crystals offers too abstract and cumbersome an approach to
categorizing these materials, and makes distinctions at too fundamental a
level between the crystallography of modulated crystals on the one hand, and
quasicrystals or intergrowth crystals on the other.

3. Except for a few solid state physicists, nobody has been paying us much
attention.  As a result we keep seeking more transparent formulations, some
of which I shall describe today.

There are several reasons for not paying attention:

{\it (a) It's trivial.} One good reason for not being interested is that we
haven't done very much.  We have merely reformulated and extended the
crystallographic classification scheme: the catalog of space groups.  Our
analysis therefore sheds no additional useful light on the real problem of
crystallography, how to infer where the atoms are from the intensity of the
peaks in the diffraction pattern.  But a crystallography that focuses too
strongly on where the atoms are may be too limited in its scope. Liquid
crystals, for example, can be periodic with no order whatever in the atomic
positions, the periodicity being entirely associated with the local
molecular {\it orientation}.  Our view is that the more powerful, broad, and
transparent a grasp you have of the easy part of the problem---what you can
learn from symmetry alone---the better equipped you will be to get on with
the hard parts.

{\it (b) It's tedious.} Another good reason for not being interested is that
space group theory is traditionally very boring.  I quote from a widely used
solid state physics text:\ft{Ashcroft (1976).} ``{\it We shall have
mercifully little to say about the 230 space groups $\ldots$ .\/}'' Having a
vastly more compact way to derive and extend the space-group classification
scheme is like coming up with a more compact derivation of everything in
Gradshteyn and Rizhik.  Nobody wants to see a better way to get there.  The
results already exist and are tabulated.  It's enough to be able to use
them.  And future versions of {\it Mathematica\/} will doubtless be able to
do all that and more, liberating our energies for more stimulating pursuits.

{\it (c) It's unfamiliar.} Not a good reason.  Our approach provides a
straightforward and compact way to extract the categories of quasiperiodic
crystals with crystallographic point group symmetries, in most cases
directly from a knowledge of the corresponding crystallographic space
groups, but to appreciate the straightforwardness and compactness it is
necessary to learn to look at ordinary crystallographic space groups from
our somewhat unfamiliar perspective. While we are convinced that
conventional space groups for periodic structures are also more
transparently viewed from our point of view,\ft{The first people to
recommend this perspective on periodic crystals were A. Bienenstock and P.
P. Ewald (1962).  We have further simplified their approach.} those who have
mastered the intracicies of the {\it International Tables for
Crystallography\/}, are understandably unwilling to abandon that hard won
territory to start over again on new ground.  That is fine.  If you feel
that way, stay where you are.

But if you are a student, too young to have suffered through the
conventional approach to space groups, then pay attention.  I will show you
a simple way to think about the space-group classification scheme that (1)
makes no distinction between periodic and quasiperiodic materials and no
distinctions within the family of quasiperiodic materials, (2) is
closer to the geometric approach of the founders of the subject than to the
algebraic approach that has dominated its exposition in recent years,\ft{The
algebraic approach is, of course, extremely important in formulating the
crystallography of higher dimensional spaces, where geometric intuition is
much harder to maintain.  But since our method avoids the detour
conventional crystallography makes into superspace when faced with
quasiperiodic materials, there is no reason not to take advantage of our
sense of 3-dimensional geometry.} and (3) can be motivated by asking a
concrete question in physics.

\bigskip
\centerline{{\bf II.  Crystallography as a Problem in Physics}}
\nobreak
\medskip
\nobreak The characteristic feature of periodic or quasiperiodic materials
(hereafter referred to as ``crystals'' in the more general sense of the term
that is now coming into widespread use) is that their diffraction patterns
consist of a finite number of sharp Bragg peaks.  Such diffraction patterns
indicate a material whose density\ft{In liquid crystals the periodicity may
be associated only with tensor properties; this has no effect on the
crystallographic categories, which emerge in essentially the same way as
they do for a scalar order parameter, though the extinction rules can be
different.  Here we consider only the scalar case.} is a superposition of
plane waves having amplitudes $\rhok$ with the property that if $\rho_0$ is
the largest amplitude, then only a finite number of amplitudes exceed an
arbitrarily small fraction $f$ of $\rho_0$.  The number of Bragg peaks is
finite because detection is limited by the resolution of the apparatus,
which sets the size of $f$ for a peak associated with a wave-vector \kv\ to
be observed.  As the resolution improves more peaks appear.  In periodic
systems they appear at larger and larger wave-vectors and in quasiperiodic
systems, additionally, between already observed peaks.

Crystallography, as a unified theory of both periodic and quasiperiodic
crystals, emer\-ges as a branch of physics when one asks whether there are
any restrictions that a countable set\ft{$L$ stands for $Laue$, but another
relevant $L$-word will emerge below.} $L$ of wave-vectors \kv\ must obey, if
it is to specify the plane waves appearing with non-zero coefficients in the
Fourier decomposition $$\rho(\rv) =
\sum_{\kv\in L}\rho(\kv)e^{i\kv\cdot\rv}\equ(rho)$$ of the density of a {\it
physically stable\/} material.  Since $\rho(\rv)$ is real, we must, of
course, have $$\rho(\kv)^* = \rho(-\kv),\equ(real)$$ so that if \kv\ is in
$L$ so is $-\kv$.  The question is whether any further conditions on the
structure of $L$ follow from considerations of stability.

By ``physically stable'' we mean that the density should be a local minimum
of the Gibbs free energy.  Since we are addressing a question of general
principle rather than calculating numerical parameters that characterize a
particular structure (interparticle distances, {\it etc.\/}), it suffices to
take that free energy to have the very general form introduced by
L.~D.~Landau in the 1930's, and subsequently used to make sense of
structural phase transitions in crystals.  In Landau theory the liquid-solid
transition is characterized by the appearance of an ``order parameter''
consisting of density Fourier components at non-zero wave-vectors.\ft{In the
liquid phase the density is uniform---only the wave-vector {\bf 0} appears;
the diffraction pattern consists only of the ``diffuse background''---there
are no Bragg peaks.} A material can be crystalline (periodic or
quasiperiodic) at temperaures $T$ and pressures $P$ at which the free
energy\ft{More precisely, the difference between the free energy of a
uniform liquid and a possible ordered phase} $$F =
\sum_{n=2}^\infty
\sum_{\hskip0.1truein\kv_1\ldots\kv_n}\An\rk1\cdots\rk{n}\equ(F)$$ can be made
negative for
suitable choices of the density Fourier coefficients $\rhok$; the stable
phase is characterized by the $\rhok$ that minimize $F$, metastable phases
are associated with local minima, and values of $\rhok$ that are not
stationary points of $F$ characterize physically unstable structures, for
which there is no free energy barrier against deforming into more favorable
configurations.  The coefficients $A$ appearing in this expansion depend on
the temperature and pressure, and are restricted in form only by the
requirement that they have the symmetries of the isotropic translationally
invariant liquid.  Translational invariance requires $\rho'(\rv) =
\rho(\rv+\dv)$ to have the same free energy as $\rho(\rv)$ for arbitrary
displacements \dv.  Since the corresponding Fourier coefficients are related
by $$\rho'(\kv) = e^{i\kv\cdot\dv}\rhok,\equ(xgauge)$$ this restricts the
non-vanishing coefficients in \(rho) to sets of wave-vectors that add up to
zero,
$$A(\kv_1,\ldots,\kv_n)= 0\ \ {\rm unless\/}\ \
\kv_1+\cdots+\kv_n=0.\equ(trans)$$ In addition rotational invariance
constrains the coefficients by the requirement $$
\A{g}n = \An \equ(rot)$$ where $g$ is any (proper or improper) 3-dimensional
rotation.  Any relations among the coefficients $A$ beyond these (or any
vanishings of individual coefficients) are accidents, removable by slight
variations in the ambient temperature and pressure.  If a wave-vector is
absent from the Fourier expansion \(rho) only as a result of such accidents,
we refer to its absence as ``accidental'' or ``non-generic''.  A ``generic''
density is one with no accidentally zero Fourier coefficients.

Landau's expansion has been widely used to investigate the possible phase
transitions between periodic structures of specified types, or, more
recently, transitions between periodic and quasiperiodic structures.  Here,
we put it to a new and more fundamental use, asking the extent to which it
limits the full range of all possible generalized crystalline
structures.

To answer this question we must first note an important consequence of the
condition
\(trans). If densities $\rho$ and $\rho'$ are related by $$\rho'(\kv) =
e^{2\pi i\chi(\kv)}\rhok \equ(gauge)$$ where $\chi$ is linear\ft{Linearity
in this context means only that $\chi(\kv_1-\kv_2) = \chi(\kv_1) -
\chi(\kv_2)$, whenever $\kv_1$, $\kv_2$, and $\kv_1 - \kv_2$ are all in
$L$.  Note that since $L$ does not contain all wave-vectors, the condition
of linearity is in general less restrictive than it is on functions of a
continuous variable
\kv.} to within an additive integer on the set $L$ of wave-vectors at which
$\rho(\kv)\neq 0$, then \(trans) requires $\rho$ and $\rho'$ to have
identical free energies.\ft{One easily shows the converse:  that any two
densities, whose Fourier coefficients have identical products over all sets
of wave-vectors summing to zero, must be related by \(gauge), so \(gauge) is
also the most general constraint translational invariance alone imposes on
degenerate densities.} If the crystal happens to be periodic, so that the
vectors in $L$ are the integral linear combinations of three generating
vectors $\b1$, $\b2$, and $\b3$ then one easily establishes that $\chi(\kv)$
is of the form $\kv\cdot\dv$, so that $\rho$ and $\rho'$ differ only by a
translation.\ft{The required $\dv$ is just $\sum\chi(\bi)\ai$, where the
$\ai$ are three vectors dual to the $\bi$ satisfying $\ai\cdot\bj =
2\pi\delta_{ij}$.  One must enter superspace if one wishes to construct a
set of $\ai$ when $D$ vectors $\bi$ are needed to generate $L$, with $D$,
the {\it rank\/} of $L$, greater than 3.  Such $\ai$ are only needed,
however, if one is wedded to periodicity as the starting point for
constructing the crystallographic categories.} If the set $L$ is generated
by more than three vectors, $\rho$ and $\rho'$ will be more generally
related.\ft{They are then said to differ by a {\it phason\/} as well as a
translation.} Since the degeneracy of their free energies is generic, it
must be that any local configuration found in $\rho(\rv)$ must also be found
in $\rho(\rv')$ with the same rate of occurrence (or one could find a
translationally invariant free energy that would discriminate between the
two).  The two densities can therefore truely be called {\it
indistinguishable} even though they differ by more than just a translation,
since it is impossible to tell whether any finite specimen is a chunk of
$\rho$ or of $\rho'$.  We call the transformation
\(gauge) between two indistinguishable densities a {\it gauge
transformation} in obvious analogy to the gauge transformations of
electrodynamics, which change the phase of a quantum mechanical
wave-function without altering any measurable properties, and we call the
function $\chi$ a {\it gauge function}.

We now address the question of interest: If $L$ is a set of wave-vectors at
which a generic density \(rho) has non-vanishing Fourier coefficients, what
features are imposed on the structure of $L$ by the requirement that the
density $\rho$ be a local minimum of the free energy $F\,$?  As noted above,
$L$ necessarily contains the negative of each of its vectors.  Less
trivially, if a Fourier coefficient $\rho(\kv_n)$ can couple {\it linearly}
in $F$ to a non-vanishing collection of terms built out of coefficients $A$
and non-vanishing Fourier coefficients
$\rho(\kv_1),\ldots\,\rho(\kv_{n-1})$, then $\rho(\kv_n)$ must itself be
non-vanishing, since otherwise the free energy would not be minimum: it
could be lowered by the appearance of a small non-zero $\rho(\kv_n)$ with
the appropriate phase to make the change in $F$ negative.  It is the absence
of such linear instabilities in a density minimizing $F$, that determines
the generic structure of the set of wave-vectors $L$.

Because the free energy coefficients $\An$ vanish unless
$\kv_1+\cdots+\kv_n=0,$ the only wave-vectors for which such linear
instabilities are possible are integral linear combinations of vectors in
$L$.  We call the set of all integral linear combinations of the vectors in
any set $L$ the {\it lattice generated\/} by $L$, and define the {\it rank}
(or {\it indexing dimension\/}) of $L$ to be the smallest number of vectors
that generate that lattice.\ft{In the superspace treatment of quasiperiodic
crystals the term ``lattice'' is only used if the rank of the lattice
generated by $L$ does not exceed the dimensionality of the wave-vectors;
when it does the term ``$Z$-module'' is used instead.  We prefer to retain
the term ``lattice'' in the general case, since the lattice plays a role
completely analysis to the (reciprocal) lattice of periodic crystals, and it
sounds absurd to say, for example, that an {\it Umklapp\/} process is one in
which wave-vector is conserved to within an additive vector of the
$Z$-module.  Our terminology sometimes requires us to be more explicit when
singling out a particular subset of the lattice generated by $L$ that is a
lattice even in the narrower sense of the term, but since there are, in
general, many such sublattices of interest, this is not such a bad thing.}
Suppose then that $\kv_1\ldots\kv_{n-1}$ are all in $L$---{\it i.e.\/}
$\rk1\ldots\rho(\kv_{n-1})$ are all non-zero---but that $\kv_n$ is not, with
$\kv_n = -\kv_1-\cdots-\kv_{n-1}$.  A generic free energy $F$ will then
certainly contain non-zero terms in which $\rk{n}$ appears linearly, and
therefore $\rk1$ can indeed be zero only if there are generic reasons for
the sum of all such terms to vanish.  In the absence of such generic
reasons, we can conclude that any integral linear combination of vectors in
$L$ is also in $L$---{\it i.e.\/} that $L$ is itself a lattice.

To complete the characterization of the general form for $L$ we must
therefore inquire into the conditions under which all terms that couple
linearly to $\rk{n}$ can vanish generically, even though $\kv_n$ is in the
lattice generated by vectors at which the density Fourier coefficients are
non-zero.  Since the only other generic property of the free energy
coefficients $A$ is the condition
\(rot) of rotational invariance, we cannot expect to find generic relations
among different terms in which $\rk{n}$ appears linearly, unless $L$ itself is
invariant under some subgroup $G_L$ of the full rotation group $O(3)$.  If
$L$ has such a symmetry and if $\kv_n$ is invariant under all the operations
in some subgroup $H$ of $G_L$, then any term that couples linearly to $\rk{n}$
will occur among a collection of terms of the form $$\sum_{g\ep
H}\A{g}n\rho(g\kv_1)\cdots\rho(g\kv_n) =
  \An\rk{n}\sum_{g\ep H}\rho(g\kv_1)\cdots\rho(g\kv_{n-1}).\equ(linear)$$
Such a collection of terms could indeed vanish, independent of the structure
of the Landau coefficients, if the density Fourier coefficients at
wave-vectors related by operations from $H$ were appropriately related, so
we must inquire into possible symmetries of the density.

We define the point group $G$ of the density as the set of operations from
$O(3)$ that take the density into an {\it indistinguishable\/} density.
Since $\rhok$ is only defined on $L$, $G$ is a subgroup of $G_L$.  In the
case of periodic crystals indistinguishable densities can differ only by a
translation, so a point group operation is, familiarly, one which leaves the
density invariant, possibly in combination with a suitable translation.  In
the general case, however, our definition of the point group is
unconventional,\ft{It is mathematically equivalent to the definition used in
the superspace approach; what is unconventional is our defining the point
group directly in terms of indistinguishability in ordinary 3-dimensional
space.} though physically compelling.  Using the Fourier space condition
\(gauge) for two densities to be indistinguishable, we note that $g$ is a
point group operation if $$\rho(g\kv)= e^{2\pi
i\P{g}\kv}\rho(\kv),\equ(phase)$$ where the gauge function $\Phi_g$, known
in this context as a {\it phase function\/}, is linear (modulo an additive
integer) on $L$.  If we take $H$ to be a non-trivial subgroup\ft{If the
group $H$ leaving $\kv_n$ invariant has only the identity in common with $G$
then there is no generic reason for $\kv_n$ to be missing from $L$} of $G$,
then we can remove further non-generic structure from the summation in
\(linear) to reduce it to $$
\An\rk1\cdots\rk{n}\sum_{g\ep H}e^{2\pi
  i[\P{g}{\kv_1}+\cdots+\P{g}{\kv_{n-1}}]}. \equ(linphsum)$$ Because $\Phi_g$
is
linear on $L$, it has a unique extension to the lattice generated by $L$,
and therefore to $\kv_n$.  Since $\kv_n$ = $-\kv_1-\cdots-\kv_{n-1}$, the
sum on the right side of
\(linphsum) is just $$\sum_{g\ep H}e^{-2\pi i\P{g}{\kv_n}}.\equ(linph)$$

Are there conditions under which this sum must vanish?   Since the
sum is over a subgroup $H$, if $h$ is any element of $H$ the sum can equally
well be written as $$\sum_{g\ep H}e^{-2\pi i\P{gh}{\kv_n}}.\equ(linh)$$
But as a direct consequence of the definition \(phase), the phase function
associated with the product $gh$ of any two operations from $G$ is related
to those associated with the operations themselves by the {\it group
compatibility condition\/} $$\P{gh}\kv \=
\P{g}{h\kv} + \P{h}\kv,\equ(gcc)$$ where we use $\equiv$ to denote equality
modulo unity.  Applying this to the sum in \(linh), and noting that every
operation $h$ in $H$ leaves $\kv_n$ invariant, we see that the sum reduces
back to its original form \(linph), except for the phase factor $e^{-2\pi
i\P{h}{\kv_n}}$.  Therefore the sum must indeed vanish unless $e^{-2\pi
i\P{h}{\kv_n}}=1$ for every $h$ in $H$ (in which case the sum is non-zero,
being simply the order of $H$).

Thus {\it a vector \kv\ from the lattice generated by $L$ can be missing
from the diffraction pattern of a generically stable density only if \kv\ is
invariant under at least one element $g$ of the point group $G$ of the
density, for which the phase function $\P{g}{\kv}$ does not vanish modulo
unity.\/}\ft{Note that the {\it converse} of this proposition is both
trivial and familiar in the periodic case: if $g\kv = \kv$ but $\P{g}{\kv}$
is not integral, then \(phase) requires $\rhok$ to vanish.} In the periodic
case the lattice is generated by just three vectors $\bi$ and the phase
function $2\pi\P{g}{\kv}$ reduces to $\kv\cdot\av_g$, where $\av_g$ is any
translation that combined with point group operation $g$ leaves the density
{\it identical\/}. If $\kv\cdot\av_g$ is non-integral, then $\av_g$ is not a
vector of the direct lattice spanned by the $\ai$ dual to the $\bi$, and one
recovers the usual rules for {\it extinctions\/} in periodic crystals.

What is novel about this formulation is (a) that it demonstrates as a
condition of stability that this is the {\it only} way in which a set of
wave-vectors with non-vanishing Fourier coefficients can fail to be a
lattice, except for accidental vanishings at isolated points in the phase
diagram and (b) that it establishes even in quasiperiodic crystals the fact
(which is not at all obvious from the superspace point of view) that the only
structures for $L$ that can characterize a physically stable 3-dimensional
quasiperiodic crystal continue to be lattices or lattices thinned by the
above generalization of the extinction condition for periodic
crystals.\ft{The result is not without its subtleties even in the periodic
case.  A hexagonal close packing of spherically symmetrical atoms, for
example has its non-zero density Fourier coefficients on a set $S_0$ that is
{\it not\/} a lattice thinned by the extinctions of a space group---there
are too many missing wave-vectors.  Only when one remembers that generic
interactions will deform each atom to the symmetry of its environment in the
crystal, do the ``missing'' points appear.  If the coupling of an atom to
its neighbors is weak, the corresponding Bragg peaks may well be very faint,
but they are there in principle, and crystallography does not give the set
$S_0$ the fundamental status it assigns to lattices or lattices with
extinctions.}

It is therefore necessary to understand the different sets of phase functions
one can associate with the subgroups\ft{The only subgroups one need consider
are those to which the symmetry of the lattice cannot be reduced by an
infinitessimal distortion.  It should be possible to prove this by a Landau
theoretic argument of generic instability against such lattice distortions.}
of the lattice point group $G_L$. The phase functions can be grouped into
equivalence classes according to two rules:

1. Phase functions that can characterize indistinguishable densities should
clearly be placed in the same class.  Comparing the condition \(gauge) that
two densities be indistinguishable with the definition \(phase) of the phase
functions, we conclude that two sets of phase functions $\Phi$ and $\Phi'$
are in the same class if there is a function $\chi$ linear (modulo 1) on the
lattice $L$ such that $$\Phi_g'(\kv) \= \Phi_g(\kv) +
\chi([g-1]\kv)\equ(gequiv)$$ for all $g$ in $G$.  Phase functions related by
\(gequiv) are called {\it gauge equivalent\/} and \(gequiv) is referred to
as a {\it gauge transformation\/} of the phase functions.\ft{In the periodic
case gauge transformations correspond to a change of origin.} Note that the
gauge transformation \(gequiv) does not alter the value of a phase function
at any vector left invariant by the associated point group element, so the
condition for extinctions is gauge invariant.

2.  Two distinct classes of type 1 should be identified under the following
circumstances: Suppose there is a linear operation $h$ on $L$ that leaves
both $L$ and the point group $G$ of the material invariant:  $hGh^{-1} =
G$.  Then materials characterized by phase functions $\Phi'_g(\kv) =
\Phi_{hgh^{-1}}(h\kv)$ cannot be sensibly distinguished from those
classified by phase functions $\Phi_g(\kv)$ so that $\Phi'$ and $\Phi$
should be grouped in the same class.  If $h$ is itself an operation of $G$,
then it is a simple exercise to show from \(gcc) that $\Phi'$ and $\Phi$ are
gauge equivalent and therefore already in the same class as a consequence of
rule 1, but if $h$ is not an element of $G$, the two classes should continue
to be identified.\ft{If $h$ has a positive determinant one can, in fact,
smoothly interpolate between the two materials without at any step altering
either the point group $G$ or the rank of $L$.  If $h$ has a negative
determinant then one cannot, and there are grounds for preserving the
distinction.  Whether or not one does corresponds in the periodic case to
whether or not one distinguishes between enantiamorphic pairs of space
groups, counting 230 or 219 distinct types.} In the periodic case $h$ can be
an element of $O(3)$ (for example a 90 degree rotation when $G$ is a
tetrahedral point group, or an element of $O(3)$ combined with a rescaling
of the lattice primitive vectors (for example 90 degree rotations of an
orthorhombic lattice.) In the quasiperiodic case it can be a pure
rescaling.  For this reason we characterize two classes of phase functions
equivalent under this second rule as {\it scale equivalent\/}.

The distinct classes of phase functions under gauge equivalence and scale
equivalence correspond precisely to the space groups in the periodic case,
and constitute the generalization of the space-group classification scheme
to the general case.

What can we say about extinctions in the general quasiperiodic case?  Proper
rotations in 3-dimensions leave invariant one-dimensional subspaces---the
rotation axes.  The only improper point group operations with non-trivial
invariant subspaces are mirrors, which have invariant planes.  If $m$ is a
mirror then $m^2$ is the identity, $e$.  Since the phase function $\Phi_e$
is integral by definition, applying
\(gcc) to $\Phi_{m^2}$ we learn that if \kv\ is invariant under $m$ then
$2\Phi_m(\kv)$ is integral, so $$\Phi_m(\kv) \= 0\ {\rm or}\
\fr1/2.\equ(glide)$$ If $\Phi_m$ is not 0 for all \kv, we have the
generalization to quasiperiodic materials of a {\it glide-plane}.  Applying the
same argument to the identity $r^n = e$ obeyed by an $n$-fold rotation $r$,
we learn that if
\kv\ is on the axis of $r$ then $n\Phi_r(\kv)$ is integral, so $$\Phi_r(\kv)
\= 0, \fr1/n, \fr2/n,\ldots, \fr{n-1}/n. \equ(screw)$$ If any non-zero
values occur we have the generalization of a {\it screw-axis}.  When both
types of operations are present the extinctions may be further constrained.
If, for example, we have a rotation $r$ and a mirror $m$ that leaves the
axis of $r$ invariant, then $m'$ = $mr$ is another such mirror.  Applying the
condition \(glide) to $\Phi_{mr}(\kv)$ and taking $\kv$ to be along the axis
of $r$, we find from \(gcc) that $$\Phi_m(\kv) + \Phi_r(\kv) \= 0\ {\rm or}\
\fr1/2 \equ(rm)$$ which in conjunction with \(glide) and \(screw) prohibits
screw axes lying in mirror planes when $n$ is odd, and restricts them to
being of order $\fr n/2$ when $n$ is even.\ft{Although these results are
geometrically obvious in periodic crystals, we are deriving them here in the
general quasiperiodic case, even when the diffraction pattern cannot be
finitely indexed (in which case the superspace approach would have to view
the crystal as a slice of a crystal periodic in infinitely many dimensions.)}

When the lattice of wave-vectors determined by the diffraction pattern has
finite rank then the procedure for determining the phase functions is
straightforward, because they need be specified only by their values at the
lattice generating vectors, and only for a set of elements $g$ sufficient to
generate the point group $G$. (The procedure is less straightforward if one
uses ``non-primitive generating vectors''---vectors that are not themselves
in the lattice.)  These values are constrained by applying
\(gcc) to the point group generating relations.  By judiciously exploiting
the freedom to alter the phases by a gauge transformation \(gequiv),
one can simplify the calculation at the start by setting many of the unknown
phases to zero, arriving in this way at a unique representative of each
class of phase functions. In the rest of this lecture I will describe some
results of this procedure.

\bigskip
\centerline{{\bf III. Example: Crystals of Trigonal Type.}}
\medskip
\nobreak As an example I describe some classes of phase functions that are
quasiperiodic generalizations of the phase functions of periodic crystals in
the trigonal system.  Part A describes how phase functions compactly
summarize the conventional classification of the trigonal system into 25
space groups in the periodic case.  Part B describes how the phase functions
of part A generalize to crystals that are quasiperiodic in a (horizontal)
plane but periodic in the orthogonal (vertical direction).  For these axial
quasicrystals the trigonal crystal system generalizes to the {\it trigonal
quasicrystal type\/}, whenever there is a unique $n$-fold axis with $n$ any
power of an odd prime.  The 25 trigonal space groups become a particular
example ($n=p=3$) of the $3n + 2n/p + 14$ classes of phase functions.  Part
C describes an alternative generalization to crystals that are periodic in
the horizontal plane but quasiperiodic in the vertical direction.  Here
$n$-remains 3, but in the simplest case the rank of the lattice increases
from 3 to 4, giving a set of phase functions that are very simply related to
the classes in the periodic case.  When one further specializes to weakly
modulated crystals, the 49 (3+1)-superspace groups in the trigonal system,
are straightforwardly recovered from these phase functions.  If refrains
from making distinctions that are only relevant to weakly modulated
crystals, then the phase functions give only 24 distinct space-group
categories, whose different ``settings'' correspond to the 49 superspace
groups.

\bigskip
\centerline{{\sl III.A. Phase functions for Periodic Crystals}}
\nobreak\centerline{{\sl  of the Trigonal System.}}
\nobreak\medskip
\nobreak
There are just two lattices compatible with a unique 3-fold axis: the
primitive $P$-lattice, obtained by a vertical stacking of horizontal
2-dimensional triangular lattices directly on top of each other (``vertical
stacking''), and the rhombohedral $R$-lattice, obtained by stacking them
with a horizontal shift by 1/3 of a suitable vector in the 2-dimensional
horizontal sublattice (``staggered stacking'').\ft{The shift can be taken to
be by \ab, where \av\ and \bv\ are two vectors of equal magnitude 120 degrees
apart, that generate the horizontal sublattice.} There are 5 trigonal point
groups, $\bar3,\ 3,\ \bar3m,\ 3m,\ $ and 32.  The vertical mirror ``$m$'' and
horizontal 2-fold axis ``2'' have a unique orientation in the $R$ lattice,
but can be oriented in two distinct ways with respect to the $P$ lattice, as
indicated by the bifurcation in notation: $\bar3m \rightarrow \bar3m1,\
\bar31m;$ $3m \rightarrow \ 3m1,\ 31m;$ and $32 \rightarrow \ 321,\ 312$. The
international tables list 25 trigonal space-groups, 7 on the $R$-lattice and
18 on the $P$-lattice.  We can specify these in terms of the phase functions
associated with the point group generators $\bar r_3\ (\bar3),\ r_3\ (3),\
r_2\ (2)$, or $m$ that appear in their names, as follows:

\topinsert
\centerline{{\bf Phase Functions for }}
\centerline{{\bf the Periodic Trigonal System}}
\centerline{{\bf ($n=3$)}}

\bigskip
\centerline{
\setbox\strutbox=\hbox{\vrule height15pt depth9.5pt width0pt}
\vbox{\offinterlineskip
\hrule
\halign{\vrule\hfil\tt #\thinspace\hfil
        &\vrule\vrule\vrule\hfil\tt #\thinspace\hfil
        &\vrule\hfil\tt #\thinspace\hfil
        &\vrule\hfil\tt # \hfil\vrule\cr
\strut $R$ & $\bar3$ &   3 \hp 32 & $\bar3m$ \hp $3m$   \cr
\noalign{\hrule}
\strut    &  ---   &          ---                    & $\Phi_m(\cvs)$ =
$\fr1/2$ \cr
\noalign{\hrule}
\noalign{\hrule}
\noalign{\hrule}
\noalign{\hrule}
&&&$\phantom{e}$\cr
\strut $P$ & $\bar3$ & 3 \hp $\matrix{321\cr312}$ &
     $\matrix{\bar3m1\cr\bar31m}$\hp $\matrix{3m1\cr 31m}$ \cr
\noalign{\hrule}
\strut    &   ---   & $\Phi_r(\cv) = \fr1/3,\tt \fr2/3$ & $\Phi_m(\cv)$ =
$\fr1/2$\cr
\noalign{\hrule}}}}
\medskip
\narrower
\noindent{\bf Table 1.} A complete specification of the 25 trigonal space
groups for periodic crystals.  It suffices to give all non-zero phases at
lattice generating vectors for the point group generators specified by the
international symbol identifying the point group. For the $P$ lattice the
three point groups with 2-fold generators can have two distinct
orientations, distinguished by expanding the name $3m$ to the pair of names
$3m1$, $31m$, {\it etc.} The phase functions are given in a gauge in which
they vanish on the horizontal sublattice, so it suffices to specify their
possible non-zero values at the single out-of-plane lattice generating
vector $\cv$ ($P$-lattice) or $\cvs$ ($R$-lattice).
\vskip0.1truein
\hrule
\endinsert

1.  One can show that there is a gauge in which all phase functions vanish
on the horizontal sublattice.  Therefore the phase functions are completely
determined by their values at the single out-of-plane generating vector
which is perpendicular to the plane for the $P$ lattice (\cv) and has an
additional in-plane component for the $R$ lattice ($\cvs$, $s$ standing for
``staggered''.)

2.  If $G$ contains a vertical mirror $m$ then either all phase functions
are zero or only the phase function associated with the generator $m$ is
non-zero, with the value $\Phi_m(\cv)$ or $\Phi_m(\cvs) \= \fr1/2,$
signifying a glide plane.

3.  If $G$ has no vertical mirror, then all generators have zero phase
functions unless $G$ contains the proper rotation $r$ and the lattice is $P$,
in which case one can have $\Phi_r(\cv) = \fr1/3$ or $\fr2/3$, signifying a
3-fold screw axis.

This information is summarized in Table 1, which lists all the possible
non-zero phase functions in a gauge in which they have an especially simple
form.

The seven space groups with the $R$-lattice are just the five with zero
phase functions ({\it symmorphic\/}) associated with the 5 trigonal point
groups, plus the two additional ({\it non-symmorphic\/}) ones possible for
the two point groups containing the vertical mirror $m$.  The eighteen on
the $P$-lattice are the eight symmorphic ones plus four associated with the
four point groups $31m$, $3m1$, $\bar 3m1$, and $\bar 31m$ having vertical
glide planes, and six associated with the three point groups 3 and 312, and
321 having either $3_1$ or $3_2$ axes.

\topinsert
\def\tt{\thinspace\thinspace}

\bigskip
\centerline{{\bf Phase Functions for the}}
\centerline{{\bf  Trigonal Quasicrystal Type}}
\centerline{($n = p^s$, $p$ an odd prime )}
\bigskip

\def\hp{\hskip0.2truein}
\centerline{
\setbox\strutbox=\hbox{\vrule height14pt depth8.5pt width0pt}
\vbox{\offinterlineskip
\hrule
\halign{\vrule\hfil\tt #\thinspace\hfil
        &\vrule\vrule\vrule\hfil\tt #\thinspace\hfil
        &\vrule\hfil\tt #\thinspace\hfil
        &\vrule\hfil\tt # \hfil\vrule\cr
\strut $R$ & $\bar n$ &   $n$ \hskip0.3truein $n2$ & $\bar nm$ \hp $nm$   \cr
\noalign{\hrule}
\strut     &  ---   & $\matrix{\strut s>1\ {\rm only}:\cr
\strut\Phi_r(\cvs) = \fr1/n,\ldots,\fr1/p - \fr1/n}$ & $\Phi_m(\cvs)$ =
$\fr1/2$ \cr
\noalign{\hrule}
\noalign{\hrule}
\noalign{\hrule}
\noalign{\hrule}
&&&$\phantom{e}$\cr
\strut $P$ & $\bar n$ & $n$ \hskip0.3truein $\matrix{n21\cr n12}$ &
     $\matrix{\bar n m1\cr\bar n1m}$\hp $\matrix{nm1\cr n1m}$ \cr
\noalign{\hrule}
\strut     &   ---   & $\Phi_r(\cv) = \fr1/n,\ldots, \fr{n-1}/n$ &
$\Phi_m(\cv)$ = $\fr1/2$ \cr
\noalign{\hrule}}}}
\bigskip
\narrower
\noindent{\bf Table 2.} The classes of phase functions for quasicrystals of
the trigonal type---\ie\ for the order $n$ of the preferred axis a power of
an odd prime, $n = p^s$.  Except for the appearance of non-zero phases
associated with point groups $n$ and $n2$ on the rhombohedral lattice when
$s$ exceeds 1, the structure is the same as in Table 1 for the space groups
in the periodic case.  As in Table 1, the classes are specified by listing
in an appropriate gauge the only possible non-zero phases at lattice
generating vectors.
\vskip0.1truein
\hrule
\endinsert
\bigskip
\centerline{{\sl III.B. Phase Functions for Horizontally Quasiperiodic
Crystals}}
\nobreak
\centerline{{\sl  of Trigonal Type.}}
\nobreak
\medskip
\nobreak A standard axial quasicrystal with rotational symmetry $n$ has a
point group with a unique $n$-fold axis (associated with the $n$-fold
rotation $r_n$ or the $n$-fold rotoinversion $\bar r_n = -r_n$), and a
lattice that is a stacking of horizontal lattices given by all integral
linear combinations of a star of $n$ vectors of equal magnitude spaced
$2\pi/n$ apart in angle.  It turns out (see Mermin (1990)) that the only way
to stack such horizontal lattices while preserving the $n$-fold rotational
symmetry is directly above one another ({\it vertical stacking}) except when
$n$ is a power of a prime number, in which case there is a single additional
{\it staggered stacking}.\ft{When $n=3$ these are the primitive and
rhombohedral lattices noted above.  When $n=4=2^2$ they are the primitive
and centered tetragonal lattices.} When $n$ is a power of a prime $p$ other
than 2, all the classes of phase functions fall into a single pattern which
can be computed and displayed independent of the particular value of $n$,
which we call the {\it trigonal quasicrystal type}.\ft{Axial quasicrystal
types are named after the member of the family of lowest rotational order.}
I digress to note that the other case where staggered lattices are possible,
$n=2^s$, contains axial quasicrystals of the tetragonal type, and that the
complete list of quasicrystal types is:\ft{See Rabson (1991).}

\hskip0.3truein1. {\bf Trigonal:} $n$ a power of an odd prime.

\hskip0.3truein2. {\bf Tetragonal:} $n$ a power of 2.

\hskip0.3truein3. {\bf Hexagonal:} $n$ twice a power of an odd prime.

\hskip0.3truein4. {\bf Dodecagonal:} $n$ even but not twice a power of a prime.

\hskip0.3truein5. {\bf Pentadecagonal:} $n$ odd but not a power of a prime.

Table 2 gives the generalization of the phase functions for periodic
crystals of the trigonal system, to all other standard axial quasicrystals
of the trigonal type.  Note that the pattern is the obvious generalization
of Table 1 in the periodic case, except that the absence of screw axes on
the $R$ lattice turns out to be an artifact of the fact that the only
periodic specimen of the trigonal quasicrystal type has an $n$ which is only
the {\it first\/} power of a prime.
\bigskip
\centerline{{\sl III.C. Phase Functions for Vertically Quasiperiodic
Crystals}}
\centerline{{\sl  of the Trigonal System.}}
\nobreak\medskip
\nobreak Modulated crystals or intergrowth compounds are quasiperiodic
crystals with a point group that is one of the 32 allowed for periodic
crystals.  They differ from periodic crystals because their lattice requires
more than three generating vectors.  As an illustration of the utility of
working directly with the phase functions, I show how the phase functions
for the rank-4 trigonal crystals follow directly from those in Table 1 for
periodic trigonal crystals, and how in the special case of weakly modulated
crystals, all 49 of the 3+1 superspace groups in the trigonal system follow
directly from those phase functions.\ft{This example is based on work in
progress with Ron Lifshitz.}

We first note that a lattice with a 3-fold symmetry axis and 4 primitive
generating vectors must be a superposition\ft{A superposition of two
lattices is the set of all integral linear combinations of vectors from
either---\ie\ the smallest lattice that contains them both.} of two
stackings of a single triangular lattice generated by vectors, \av\ and
\bv\ of equal length and 120 degrees apart.  Exactly as in the ordinary
crystallographic (3+0) case, each of the two out-of-plane stacking vectors
\cv\ and $\cv'$ must either be perpendicular to the horizontal plane or have
a horizontal component that can be taken to be \ab.  If both stacking vectors
are vertical, one has a $PP$ lattice, and if one is staggered and one
vertical, one has an $RP$ lattice.  A third possibility, $RR$, need not be
considered, for both stacking vectors can be taken to have the same
horizontal component, and therefore the difference of the two, which can
replace one of them as a primitive generating vector, is vertical and the
lattice continues to be of the $RP$ type.\ft{The fourth combination, $PR$
has, of course, exactly the same meaning as $RP$.}

We need the phase functions for each point group generator at the four
vectors \av, \bv, \cv, and $\cv'$ that primitively generate the lattice.  We
take $\cv'$ always to be vertical, and take \cv\ to be vertical or of the
staggered form $\cvs$ (with a horizontal component \ab.) The possible
non-zero phase functions can be read directly from the corresponding
crystallographic information in Table 1, after noting the following:

1.  The phase functions at \av, \bv, and \cv\ are constrained by all the
conditions that apply in the crystallographic case, and there is therefore a
gauge in which they are given precisely by the entries in Table 1.

2. Because the fourth vector, $\cv'$, is perpendicular to the $a$-$b$ plane,
it is either invariant or changes sign under every point group operation.
As a result applying the group compatibility condition \(gcc) to the phase
functions of any of the point group generators to $\cvp$ gives conditions
that only involve phases at $\cvp$; the phases at
\av, \bv, and \cv\ are completely decoupled.

3. The procedure to determine the phases at $\cvp$ is therefore identical to
the procedure that determined the phases at \cv\ for the 3+0 $P$-lattice,
since the constraints on $\Phi_g(\cv)$ are also independent of the phases at
the two remaining vectors in the crystallographic case.  The available
choices can therefore also be read directly from Table 1.\ft{One could also,
of course, compute them directly using results such as \(glide),
\(screw), and \(rm).}

Table 3 gives the non-zero phases for the (3+1) trigonal space groups.  {\it
The entire content of Table 3 can be read directly from the corresponding
information for the periodic (3+0) case in Table 1.} The entries in the upper
half of each box are identical to those in Table 1.  The entries in the
lower half simply repeat (for both the $PP$ and $RP$ lattices) the
additional possible non-zero phase for $\cvp$ taken from the entries in
Table 1 for the $P$-lattice.  {\it Table 3 contains all information needed to
construct the space-group categories for any diffraction pattern with
trigonal symmetry that can be indexed by 4 vectors, whether that pattern
arises from a weakly modulated crystal, an intergrowth compound, or a more
general quasiperiodic structure.} How one uses this information to construct
the space-group categories depends on which features of the system at hand
one wishes to emphasize.\ft{This is analogous to the freedom one has in the
periodic case to represent, for example, the orthorhombic space groups in
various ``settings.''}

\topinsert
\centerline{{\bf Phase Functions for the}}
\centerline{{\bf (3+1) Trigonal System}}
\centerline{{\bf ($n$=3)}}
\bigskip
\centerline{
\setbox\strutbox=\hbox{\vrule height15pt depth9.5pt width0pt}
\vbox{\offinterlineskip
\hrule
\halign{\vrule\hfil\tt #\thinspace\hfil
        &\vrule\vrule\vrule\hfil\tt #\thinspace\hfil
        &\vrule\hfil\tt #\thinspace\hfil
        &\vrule\hfil\tt # \hfil\vrule\cr
\strut $RP$ & $\bar3$ &   3 \hp 32 & $\bar3m$ \hp $3m$   \cr
\noalign{\hrule}
\strut  &  ---   & $\matrix{\strut\cr\strut\Phi_r(\cvp) =  \fr1/3, \fr2/3}$ &
     $\matrix{\strut\Phi_m(\cvs) = \fr1/2\cr
              \strut\Phi_m(\cvp) = \fr1/2}$ \cr
\noalign{\hrule}
\noalign{\hrule}
\noalign{\hrule}
\noalign{\hrule}
&&&$\phantom{e}$\cr
\strut $PP$ & $\bar3$ & 3 \hp $\matrix{321\cr312}$ &
     $\matrix{\bar3m1\cr\bar31m}$\hp $\matrix{3m1\cr 31m}$ \cr
\noalign{\hrule}
\strut  &   ---   & $\matrix{\strut\Phi_r(\cv) = \fr1/3,\tt \fr2/3\cr
                         \strut\Phi_r(\cvp) = \fr1/3,\tt \fr2/3}$
              & $\matrix{\strut\Phi_m(\cv) = \fr1/2\cr
                         \strut\Phi_m(\cv') = \fr1/2}$\cr
\noalign{\hrule}}}}
\medskip
\narrower
\noindent{\bf Table 3.} Phase functions for (3+1) quasiperiodic crystals of
the trigonal system.  The table is constructed directly from Table 1 for the
periodic (3+0) trigonal system.  The entries for \cv\ and $\cvs$ are
identical to those in Table 1, and the entries for the vertical vector
$\cvp$ for both the $PP$ and $RP$ lattices, are identical to the phases given
for the $P$ lattice for
\cv\ in Table 1. In the special case of weakly modulated crystals, these
phase functions lead directly to the 49 (3+1) superspace groups, which can
be more generally viewed as different settings of the 24 categories that
emerge when does not choose to single out a particular (3+0) sublattice for
special attention.

\vskip0.1truein
\hrule
\endinsert

The primary application to date has been to weakly modulated crystals.  Here
it is useful to distinguish the 3-dimensional sublattice of $L$ that
contains the wave-vectors of the original unmodulated crystal, known as the
{\it lattice of main reflections\/}.  Lattice vectors not in this lattice of
main reflections are associated with {\it satellite\/} peaks, which are of
weaker intensity than the strongest main reflections, when the amplitude of
the modulation is small compared to the lattice constant of the undistorted
crystal.  All 49 of the tabulated (3+1) superspace groups for weakly
modulated crystals in the trigonal system,\ft{See Janner (1992), pps.
822-823.} can be read straightforwardly from Table 3, if one notes that by
focusing on weakly modulated crystals, the conventional scheme introduces an
asymmetry between the two stacking vectors.  One is conventionally always
taken from the lattice of main reflections, as a result of which its
vertical component is uniquely determined.  The other, called the modulation
wave-vector, characterizes the satellites, and is only determined modulo the
first.\ft{In classifying a more general rank-4 diffraction pattern with
trigonal symmetry, one is free to replace {\it both\/} stacking vectors with
their equivalent integral linear combinations given by any 2$\times$2
unimodular matrix of integers.}

On the $PP$ lattice (conventionally given Bravais class No. 24) the
conventional scheme lists 25 superspace groups, 8 of them symmorphic,
associated with the 8 point groups.\ft{As noted earlier, we count the same
point group twice when it can occur in two distinct orientations.} We can
read off the non-symmorphic superspace groups from Table 3 as follows:

When the point groups are 3, 321, or 312, and the phase function
$\Phi_r(\cv)$ associated with the main reflections is non-zero, then one can
always redefine the modulation wave-vector $\cvp$ (by adding to it, if
necessary, \cv\ or 2\cv) so that the phase associated with the modulations,
$\Phi_r(\cv')$ is zero.  Consequently $\Phi_r(\cvp)$ can be non-zero only
when $\Phi_r(\cv)=0$.  Note also that the direction of $\cvp$ is a matter of
convention, since it can be shifted by arbitrary multiples of \cv, and and
there are therefore no grounds for distinguishing between the phases
$\Phi_r(\cvp)\= \fr1/3$ and $\fr2/3$. There are thus 3 non-symmorphic
space-group categories with each of these three point groups in the
conventional scheme.  \ft{If one chooses not to distinguish between
satellites and main reflections, either because the modulations are strong
or because one is studying a more general form of quasiperiodic crystal,
then the the three non-symmorphic categories reduce to only one, since there
are then no grounds for distinguishing between phases $\Phi_r(\cv)$ =
$\fr1/3$ and $\fr2/3$, and no grounds for distinguishing between a single
non-zero phase of $1/3$ associated with \cv\ or with $\cvp$ .}

Similarly, when the point groups are $3m1$, $31m$, $\bar 3m1$, or $\bar
31m$, then the case $\Phi_m(\cvp)=\fr1/2$ arises only when $\Phi_m(\cv)\=0$,
so non-symmorphic space groups correspond to one or the other but not both
being $\fr1/2$.  Thes gives the 8 remaining non-symmorphic superspace
groups.\ft{They reduce to 4 if one makes no distinction between satellites
and main reflections.  The total of 25 superspace groups on the $PP$ lattice
thus reduces to 15 if one does not make distinctions appropriate only in
the weakly modulated case.}

Turning to the $RP$ lattice, note first that the conventional scheme for
weakly modulated crystals assigns it to two different Bravais classes, Nos.
22 and 23, depending on whether the lattice of main reflections is
rhombohedral or primitive. Therefore the information in row $RP$ of Table 3
must be cast into two distinct forms to recover all the conventional
superspace groups.  When the lattice of main reflections is rhombohedral,
the conventional scheme lists 11 superspace groups (under Bravais class 22);
when it is primitive 13 are listed (under Bravais class 23).  In either case
the number of symmorphic superspace groups is 5, because the staggered
vertical vector fixes the orientation of the mirror planes and 2-fold axes,
whether or not it is viewed as belonging to the lattice of main reflections
or characterizing the satellite peaks.  The non-symmorphic superspace
groups, in either of the conventional descriptions can be read directly from
the entries for the $RP$ lattice in Table 3 as follows:

If the lattice of main reflections is rhombohedral then $\cv'$ is the
modulation vector.  When so regarded, its vertical direction is arbitrary,
since it can be shifted by a multiple of the vertical component of the
vector $3\cvs$ from the lattice of main reflections.  We therefore have just
one non-symmorphic superspace group for each of the point groups 3 and 32,
with $\Phi_r(\cvs)$ zero and the extinctions being associated with the
satellite wave-vector $\cv'$.  With $\cv'$ defined only to within the
vertical component of $3\cvs$, there are only two distinct choices for the
mirror phases, which give the two non-symmorphic superspace groups for each
of the two point groups\ft{These four possibilities reduce to two if one
does not distinguish satellites from main reflections, reducing these 11
superspace groups to 9 in the general case.}
 $3m$ and $\bar3m$.

If, on the other hand, the lattice of main reflections is primitive, and
$\cvs$ is the modulation vector, then because the conventional scheme fixes
$\cv'$ for each of the point groups 3 and 32, we now have two non-symmorphic
superspace groups corresponding to the two phases of $\Phi_r(\cvp)$.\ft{This
explains why the conventional scheme lists two more space groups for the
$RP$ lattice when viewing it as Bravais class No.~23, than it does when
viewing it as Bravais class No.~22.  These four non-symmorphic superspace
groups reduce to two if one makes no distinction between satellites and
lattices of main reflections.} The situation with respect to the mirror
phases remains the same, giving four more non-symmorphic superspace
groups.\ft{They reduce to two if one makes no distinctions between
satellites and lattices of main reflections.  In the general case there is,
of course, no difference between the categories corresponding to
conventional Bravais classes 22 and 23, there being only a single $RP$
lattice.}

It can thus be quite useful to recognize explicitly, as the conventional
scheme does not, that the superspace groups in the conventional Bravais
classes 22 and 23 are characterized by identical lattices and identical
phase functions.  Categories in the two cases differ only by the particular
3+0 sublattice singled out for special emphasis as the lattice of main
reflections, and by the consequences of the additional rule that the
vertical component of the out-of-plane generating vector for the lattice of
main reflections cannot be altered when applying scale equivalence to
identify categories.

Note, finally, that if the crystallographic space groups from the other
non-cubic crystal systems are specified in terms of their phase functions in
the manner of Table 1, then the superspace groups for all but three of the
conventional 24 (3+1) Bravais classes for modulated crystals can be
extracted directly from the space groups for periodic crystals in exactly
the same trivial way, by introducting a 4th set of phases that can easily be
independently determined.  Three of the conventional (3+1) Bravais classes,
corresponding to one monoclinic and one orthorhombic\ft{The one orthorhombic
lattice is conventionally represented in two distinct Bravais classes
depending on which of its 3+0 sublattices one choses to regard as the
lattice of main reflections.} lattice, cannot be represented as a
crystallographic lattice augmented by an additional generating vector that
is either invariant or reversed in sign by every point group operation.  In
those two cases one must do some additional work, but that work is
conceptually no different from the work done to extract the categories in
the periodic case.

\medskip {\it Acknowledgment:} Supported by National Science Foundation
Grant DMR8920979.

\vskip1truein
\centerline{{\bf References}}
\medskip

{
\baselineskip=14pt
\parindent=0pt
\parskip=5pt

Ashcroft, N. W. and Mermin, N. D., 1976, {\it Solid State Physics\/},
Saunders College, Philadelphia.

Bienenstock, A., and P. P. Ewald, 1962, Acta. Cryst. {\bf 15}, 1253-1261

Janssen, T., A. Janner, A. Looijenga-Vos, and P. M. de Wolff, 1992,
{International Tables for Crystallography, Volume C.\/}, ed.
A.~J.~C.~Wilson, Kluwer Academic, Dodrecht, 797-844.

Mermin, N. D., 1989, {\it XVIIth International Colloquium on Group
Theoretical Methods in Physics}, eds. Y. Saint-Aubin and L. Vinet, World
Scientific, Singapore, 103-126.

Mermin, N. D., D.~A.~Rabson, D.~S.~Rokhsar, and D.~C.~Wright, 1990, Phys. Rev.
B{41}, 10498-10502.

Mermin, N. D., 1992a, Revs. Mod. Phys. {\bf 64}, 3-49.

Mermin, N. D., 1992b, Phys. Rev. Lett. {\bf 68}, 1172-5.

Mermin, N. D., and R. Lifshitz, 1992c, Acta Cryst. A, to appear.

Rabson, D. A., N. D. Mermin, D. S. Rokhsar, and D. C. Wright, 1991,
Revs. Mod. Phys. {\bf 63}, 699-733.

}

\bye